\documentstyle[aps,prb,twocolumn,bezier,floats]{revtex} \begin{document} 
\title{
   Compressibility of electron gas
   in presence of Anderson's impurity.
}
\author{ Daniel L. Miller }
\address{ Dept. of Physics of Complex Systems,\\
          The Weizmann Institute of science,
          Rehovot, 76100 Israel                \\
          e-mail  fndaniil@wicc.weizmann.ac.il
}
\date{\today}

\twocolumn[
\maketitle
   \begin{abstract} \widetext\hsize\columnwidth\leftskip=0.10753 
   \textwidth\rightskip\leftskip\nointerlineskip\small\relax
   We employ Wiegmann's solution of the Anderson impurity model in order to
   compute the compressibility of electron gas.
   We have found that there is a pair of neighbor levels separated by
   anomalously large energy $\propto L^{-1/3}$, where $L$ is the size of the
   system. The Fermi energy crosses this pseudogap and the 
   compressibility decreases near the so-called ``symmetric'' limit. 
\end{abstract} \hspace{1.5cm}] \narrowtext

The Anderson Hamiltonian\cite{Anderson-may61} describes the localized state
interacting with electron gas
\begin{eqnarray}
  \hat {\cal H} &=& \sum_{k\sigma}
  \left\{ \epsilon_k \hat a^\dagger_{k\sigma}\hat a_{k\sigma}
  + V_k \hat a^\dagger_{k\sigma}\hat a_{d\sigma}
  + V_k^\ast \hat a^\dagger_{d\sigma} \hat a_{k\sigma}
  \right\}
\nonumber\\
  &+& \epsilon_d (\hat n_\uparrow + \hat n_\downarrow)
  + U \hat n_\uparrow\hat n_\downarrow
  \;\;\;\;\;
  \hat n_\sigma =  \hat a^\dagger_{d\sigma} \hat a_{d\sigma}
  \;.
\label{eq:model.1}
\end{eqnarray}
Here $\hat a^\dagger_{k\sigma}$ creates the electron in the band with
wave number $k$, spin $\sigma$, and energy $\epsilon_k$. The operator
$\hat a^\dagger_{d\sigma}$ creates the localized electron with energy
$\epsilon_d$. The interaction between localized electrons gives the
correlation term $U \hat n_\uparrow\hat n_\downarrow$, and $V_k\hat
a^\dagger_{k\sigma} \hat a_{d\sigma}$ describes their decay to the band.

Wiegmann have  solved this problem in one dimension\cite{Wiegmann-nov80}. At
the zero magnetic field the ground state is described by the set of
rapidities $\{\lambda_\alpha\}_{\alpha = 0}^M$. They satisfy
\begin{eqnarray}
  &&
  \pi J_\alpha + \sum_{\beta = 0}^M
  \text{arctan}(\lambda_\alpha-\lambda_\beta)  = -x(\lambda_\alpha) L
\label{eq:model.2}
\\
  && x(\lambda) = \epsilon_d + U/2  -
  \sqrt{U\Gamma(\lambda + \sqrt{\lambda^2 + 1/4})}\;,
\label{eq:model.3}
\end{eqnarray}
where $L$ is the size of the sample, $\Gamma = |V|^2$, $\hbar v_F=1$, and
therefore $\epsilon_k = k$.  Terms of the order of $1/L$ are omitted, because
we are going to discuss the properties of the electron gas only. The integer
numbers $\{J_\alpha\}_{\alpha = 0}^M$ completely describe the state of the
system with $N=2(M+1)$ particles. The energy of the system is just
\begin{equation}
   E = 2\sum_{\alpha=0}^M x(\lambda_\alpha)\;.
\label{eq:model.4}
\end{equation}
The ground state corresponds to subsequent values of $J_\alpha$; we can put
$J_\alpha = {LD\over\pi}-{3M\over 2}\ldots {LD\over\pi}+{M\over 2}$, where
$D$ is the energy of the bottom of the band. In the non-interacting case,
$U=0$, one obtains the chemical potential
$\mu\equiv{dE\over dN}={2\pi M\over L}-D$.  The compressibility of the
gas  is ${dn\over d\mu}= {1\over \pi}$, where $n=N/L$ is the density of
particles.

These results remain valid for finite $U$ if one makes use of the
following continuous approximation. Under the condition
\begin{equation}
   |\lambda_{\alpha+1}-\lambda_\alpha| \ll 1\;\;\;\;\forall \alpha
\label{eq:model.5}
\end{equation}
one can transform Eq.~(\ref{eq:model.2}) to the integral equation
\begin{equation}
   \sigma_\lambda + \tilde \sigma_\lambda
   + {1\over \pi}\int_{\lambda_0}^{\lambda_M}
   {\sigma_{\lambda'} d\lambda'
       \over
   1 + (\lambda - \lambda')^2}
   = - {1\over \pi}{dx \over d\lambda}
\label{eq:model.6}
\end{equation}
where the mean density of rapidities
$\sigma(\lambda) = {1/L \over \lambda_{\alpha+1} - \lambda_\alpha}$, and
$\sigma(\lambda) \propto \theta(\lambda - \lambda_0)$. (Step function.) We
have also introduced the auxiliary function $\tilde \sigma(\lambda) \propto
\theta(\lambda_0 - \lambda)$   in order to make Eq.~(\ref{eq:model.6}) valid
for $\lambda\in ]-\infty,\lambda_M]$.

It is convenient to integrate Eq.~(\ref{eq:model.6}) from $-\infty$ to
$\lambda_M$ and combine it with Eq.~(\ref{eq:model.2})  for $\alpha = 0$
\begin{equation}
    \pi\int_{-\infty}^{\lambda_0}
    \tilde\sigma_{\lambda} d\lambda =
    \epsilon_d + U/2 + D - 2\pi {M\over L}\;.
\label{eq:model.7}
\end{equation}
This is an equation for $\lambda_0$.  In the symmetric limit
$|\epsilon_d + U/2  + D - 2\pi {M\over L}|\ll \sqrt{U\Gamma}$ one
has\cite{Wiegmann-sep82}
\begin{equation}
   \lambda_0 \approx {1\over \pi}
   \log{E_d + U/2 + D - 2\pi M / L \over 2\sqrt{U\Gamma/\pi e}}
\label{eq:model.8}
\end{equation}
goes to $-\infty$ when $E_d + U/2 + D - 2\pi M / L $ goes to zero.

\begin{figure*}
   \unitlength=1mm
   \linethickness{0.4pt}
   \begin{picture}(140.00,100.00)
   \put(0.00,40.00){\vector(1,0){140.00}}
   \put(70.00,0.00){\vector(0,1){100.00}}
   \put(72.00,95.00){\makebox(0,0)[lc]{$x(\lambda)$}}
   \put(135.00,38.00){\makebox(0,0)[ct]{$\lambda$}}
   \put(70.00,80.00){\line(-1,0){70.00}}
   \put(72.00,80.00){\makebox(0,0)[lc]{$\epsilon_d+U/2$}}
   \bezier{696}(0.00,78.00)(103.00,58.00)(140.00,0.00)
   \put(70.00,76.00){\line(-1,0){61.00}}
   \put(9.00,76.00){\line(0,-1){36.00}}
   \put(9.00,38.00){\makebox(0,0)[ct]{$\lambda_0$}}
   \put(70.00,60.00){\line(-1,0){8.00}}
   \put(62.00,60.00){\line(0,-1){20.00}}
   \put(62.00,38.00){\makebox(0,0)[ct]{$\lambda_1$}}
   \put(72.00,68.00){\makebox(0,0)[lc]{$\Biggr\}\delta\mu$}}
   \put(110.98,35.99){\line(0,-1){3.98}}
   \put(111.00,40.00){\line(0,-1){8.00}}
   \put(111.00,32.00){\line(-1,0){41.00}}
   \put(111.00,42.00){\makebox(0,0)[cb]{$U\Gamma/8$}}
   \put(72.00,30.00){\makebox(0,0)[lt]{$\epsilon_d$}}
   \put(128.00,40.00){\line(0,-1){24.00}}
   \put(128.00,16.00){\line(-1,0){58.00}}
   \put(72.00,14.00){\makebox(0,0)[lt]{$ - D$}}
   \put(128.00,42.00){\makebox(0,0)[cb]{$\lambda_M$}}
   \end{picture}
   \caption{ The main parameters of the problem. Energy levels appear
   densely between $ - D$ and $\mu\sim\epsilon_d+U/2$ (``symmetric'' limit).
   Let us just mention ``asymmetric''
   limit, when Fermi level is near $\epsilon_d$; this is the regime of the
   formation of the local moment.
}
\label{fig:xotlambd} \end{figure*}
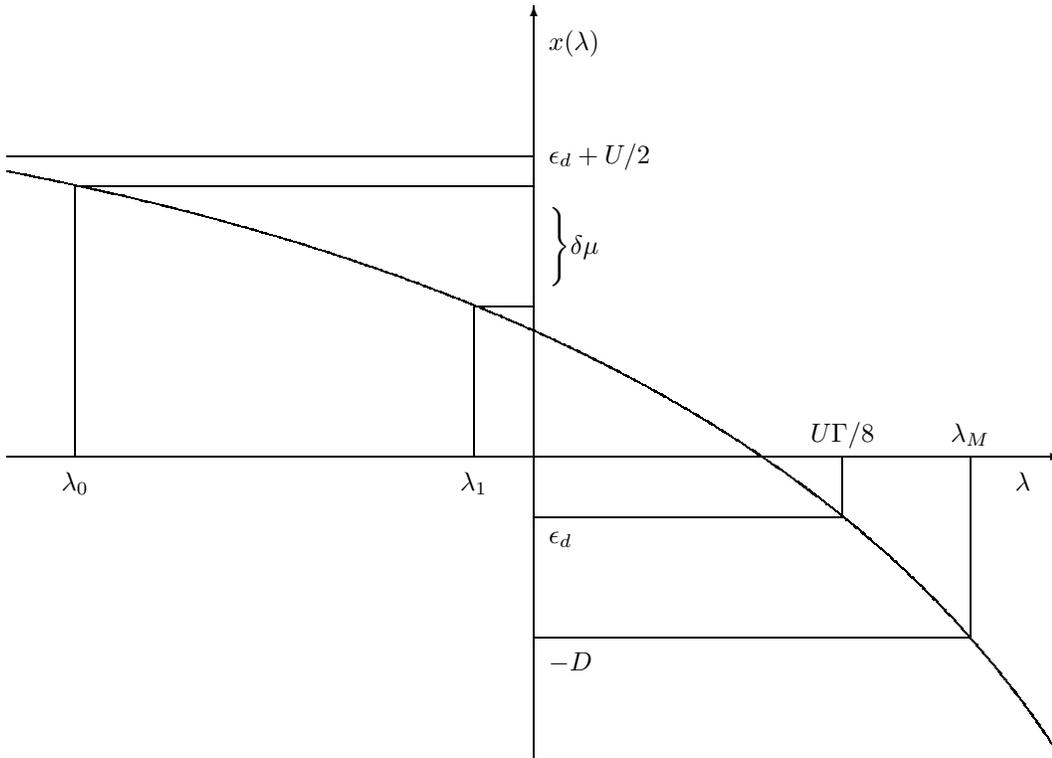

{\em The problem } is that the condition Eq.~(\ref{eq:model.5}) will
be  broken eventually. It is clear that for $\lambda_0\rightarrow-\infty$
the density of the rapidities goes to zero $\sigma(\lambda_0) <
|{dx \over d\lambda}|\sim |\lambda_0|^{-3/2}$. Therefore
\begin{equation}
     |\lambda_1-\lambda_0|  \gtrsim |\lambda_0|^{3/2}
     \rightarrow \infty \;,
\label{eq:model.9}
\end{equation}
and the solution of Eq.~(\ref{eq:model.2}) cannot be described by the
density of rapidities $\sigma(\lambda)$.

The main idea of the present paper is to introduce one rapidity $\lambda_0$
separated from all the rest $\lambda_1\ldots\lambda_M$. Then
\begin{equation}
  x(\lambda_0) = 2\pi {M\over L} - D
  \;\;\;\;\lambda_0 \approx
  -{U\Gamma/8 \over(2\pi {M/ L} - D -\epsilon_d-U/2)^2}
\label{model.10}
\end{equation}
is obtained directly from Eq.~(\ref{eq:model.2}) with $\alpha = 0$ under
the condition $\lambda_\alpha-\lambda_0 \gg 1$ for  $\alpha = 1\ldots M$.
The rest of equations gives
\begin{equation}
   \sigma_\lambda + \tilde \sigma_\lambda
   + {1\over \pi}\int_{\lambda_1}^{\lambda_M}
   {\sigma_{\lambda'} d\lambda'
       \over
   1 + (\lambda - \lambda')^2}
   = - {1\over \pi}{dx \over d\lambda}
\label{eq:model.11}
\end{equation}
The solution is well known and instead of Eq.~(\ref{eq:model.8})
we will have
\begin{equation}
   \lambda_1 \approx
   {1\over 2\pi} \log{1\over |\lambda_0|}
\label{eq:model.12}
\end{equation}
It means appearance of a gap in the excitation spectrum
and a jump of the chemical potential
\begin{equation}
   \delta\mu = 2[x(\lambda_0)-x(\lambda_1)]
   \approx \sqrt{\pi U\Gamma\over \log|\lambda_0|} \;.
\label{eq:model.14}
\end{equation}
The gap is opened when the condition
Eq.~(\ref{eq:model.5}) is broken for $\lambda_1-\lambda_0$. This occurs
when $ (L/\pi) |dx/d\lambda| \approx 1 $ or
\begin{equation}
   2\lambda_1
   \approx \left( {U\Gamma L^2 \over 4\pi^2} \right)^{1/3}
\label{eq:model.16}
\end{equation}
So our main result is derived from
Eqs.~(\ref{eq:model.12})-(\ref{eq:model.16})
\begin{equation}
   \delta \mu  \approx \left( {\pi U\Gamma \over L} \right)^{1/3}
\label{eq:model.15}
\end{equation}
and the gap is anomalously large, because it is proportional to  $L^{-1/3}$.
The compressibility is suppressed
${\delta n\over \delta \mu}\sim (U\Gamma L^2)^{-1/3}\ll 1$.

It is important to mention that the model above is valid for even number of
electrons in the system, and the computed above gap is for two  additional
particles. The energy of the intermediate state with odd number of electrons
can be computed also by the method presented here. It is necessarily to
introduce equation for the charge rapidity and to couple it to
Eq.~(\ref{eq:model.2}).  Then our huge gap will be split in two parts.
However, at least one of them must be large.

In conclusion we have found the anomalously large spacing between
energy levels of electron gas interacting with impurity. This is a 
finite size effect, which requires special treatment.
The pseudogap appears at the Fermi energy when 
the model approaches complete particle-hole symmetry,
$\mu \approx \epsilon_d+U/2$.

\acknowledgments
This work was supported by Israel Science Foundation and the Minerva Center
for Nonlinear Physics of Complex systems.

\end{document}